\documentclass[journal=jctcce,manuscript=article]{achemso}
\usepackage{dcolumn} 
\usepackage{graphicx} 
\usepackage[version=3]{mhchem}
\usepackage{color,soul}
\usepackage{multirow}
\usepackage{braket}
\usepackage{xcolor}
\usepackage{bm}
\usepackage{comment}
\usepackage{hyperref}
\usepackage{physics}   

\newcommand{\hbm}[1]{\hat{\bm{#1}}} 
\newcommand{\pp}[2]{\frac{\partial{#1}}{\partial{#2}}}
\newcommand{\bGamma}{{\bm \Gamma}}

\newcommand{\hH}{\hat H}

\newcommand{\hP}{\hat P}

\newcommand{\bR}{{\bm R}}
\newcommand{\bP}{{\bm P}}

\newcommand{\bd}{{\bm d}}

\newcommand{\cev}[1]{\reflectbox{\ensuremath{\vec{\reflectbox{\ensuremath{#1}}}}}}

\newcommand{\mycomment}[1]{{\color{black}{#1}}}

\title{A phase-space view of vibrational energies without the Born-Oppenheimer framework}
\author{Xuezhi Bian}
\affiliation{Department of Chemistry, Princeton University, Princeton, New Jersey 08544, USA}
\author{Cameron Khan}
\affiliation{Department of Chemistry, Princeton University, Princeton, New Jersey 08544, USA}
\author{Titouan Duston}
\affiliation{Department of Chemistry, Princeton University, Princeton, New Jersey 08544, USA}
\author{Jonathan Rawlinson}
\affiliation{School of Science \& Technology, Nottingham Trent University, Nottingham
NG1 4FQ, UK}
\author{Robert G. Littlejohn}
\affiliation{Department of Physics, University of California, Berkeley, California 94720, USA}
\author{Joseph E. Subotnik}
\email{subotnik@princeton.edu}
\affiliation{Department of Chemistry, Princeton University, Princeton, New Jersey 08544, USA}
\affiliation{Department of Chemistry, University of Pennsylvania, Philadelphia, Pennsylvania 19104, USA}

\date{\today}

\begin{document}
\begin{abstract}
    We show that  following the standard mantra of quantum chemistry and  diagonalizing the Born-Oppenheimer (BO) Hamiltonian $\hat H_{\rm BO}(\bR)$ is not the optimal means to construct potential energy surfaces. A better approach is to diagonalize a phase-space electronic Hamiltonian, $\hat H_{\rm PS}(\bR,\bP)$, which is parameterized by both nuclear position $\bR$ and nuclear momentum $\bP$. The foundation of such a non-perturbative phase-space electronic Hamiltonian can be made rigorous using a partial Wigner transform and the method has exactly the same cost as BO for a semiclassical calculation (and only a slight increase in cost for a quantum nuclear calculation).  For a three-particle system, with two heavy particles and one light particle, numerical results show that a phase-space electronic Hamiltonian produces not only meaningful electronic momenta (which are completely ignored by BO theory) but also far better vibrational energies. As such, for high level results and/or systems with degeneracies and spin degrees of freedom, we anticipate that future electronic structure and quantum chemistry packages will need to take as input not just the positions of the nuclei but also their momenta.
\end{abstract}

\maketitle
\section{Introduction} \label{sec:intro}
\subsection{The Hydrogen Atom}

The Born-Oppenheimer (BO) framework\cite{heitler1927,born1927,born1996} is the standard approach to understanding chemistry, upon which the vast majority of chemical intuition is built.
The basic ansatz of BO theory is that, because nuclei are slow and electrons are fast, one can separate the total Hamiltonian into two terms, the nuclear kinetic energy and the electronic Hamiltonian: 
\begin{eqnarray}\label{eq:Hgeneral}
    \hH = \frac {{\hat {\bm P}}^2} {2M} + \hat H_{\rm el}(\hat {\bm R}).
\end{eqnarray}
One diagonalizes $\hat H_{\rm el}(\hat {\bm R})$ in order to construct potential energy surfaces, and then the BO approximation stipulates that nuclear motion follows a single surface.

One critical limitation of BO theory can  be identified by considering the hydrogen atom, the most well-understood problem in quantum mechanics. The Hamiltonian is simply the kinetic energy of the electron plus the kinetic energy of the proton plus their interaction:
\begin{eqnarray} \label{eq:HHatom}
    \hat H_{\rm Hydrogen}  &=& \frac {\hat {\bm P}_{\rm n}^2} {2M} + \hat H^{\rm el}_{\rm Hydrogen} \\
    \label{eq:Helhydrogen}
     \hat H^{\rm el}_{\rm Hydrogen} & = & \frac {\hat {\bm p}_{\rm e}^2} {2m_{\rm e}}   - \frac {e^2} {4\pi \epsilon_0 |{\bm {\hat r}_{\rm e} - \bm {\hat R}_{\rm n}}|}.
\end{eqnarray}

\subsection{Standard Approach: Separate out the center of mass}

As can be found in any textbook\cite{cohen2019}, the way to solve this Hamiltonian is to change variables from $\hat{\bm R}_{\rm n}$ and $\hat{\bm r}_{\rm e}$ to the relative position and center of mass coordinates:
\begin{eqnarray}
    \hat {\bm r} =  \hat{\bm r}_{\rm e} -  \hat{\bm R}_{\rm n}, \quad  
    \hat {\bm R} = \frac {M  \hat {\bm R}_{\rm n} + m_{\rm e}  \hat{\bm r} } {M + m_{\rm e}}.
\end{eqnarray}
\mycomment{with 
the corresponding momentum operators  $\hat {\bm p}$ and $\hat {\bm P}$ in the center of mass coordinates as well:
\begin{eqnarray}
    \hat {\bm p} = \hat{\bm P_n} - \hat{\bm p_e} , \quad  \hat{\bm P} = \hat{\bm P_n} + \hat{\bm p_e}. 
\end{eqnarray}
}
In doing so, the Hamiltonian becomes:
\begin{eqnarray} \label{eq:Hcom}
    \hat H_{\rm com} = \frac {\hat {\bm P}^2} {2(M + m_{\rm e})} +\frac {\hat {\bm p}^2} {2\mu} - \frac {e^2} {4\pi\epsilon_0|\hat {\bm r}|}
\end{eqnarray}
Here,  the reduced mass is $\mu = (M^{-1}+m_e^{-1})^{-1}$. 
The stationary energies for the system are then the center of mass kinetic energy plus the quantized bound state energy of the atom:
\begin{eqnarray}\label{eq:Eexact}
    E_{n}^{\rm exact}(\bm P) = \frac{\bm P^2}{2(M + m_{\rm e})} -\frac{\mu e^4}{8 \epsilon_0^2 h^2} \frac{1}{n^2}.
\end{eqnarray}
Note that, within the BO (clamped-nuclei) approximation, the result is slightly different if one diagonalizes $\hH_{\rm el}$:
\begin{eqnarray} \label{eq:Ecn}
    E_{n}^{\rm BO}(\bm P_{\rm n}) = \frac{\bm P_{\rm n}^2}{2M } -\frac{m_{\rm e} e^4}{8 \epsilon_0^2 h^2} \frac{1}{n^2}.
\end{eqnarray}

Now, one might argue that the energy expressions are not that different between Eq.~\ref{eq:Eexact} and Eq.~\ref{eq:Ecn}. That being said, if we ignore the raw energy difference for a moment, there is a conceptual problem insofar as the fact that, unless one changes coordinates, the electronic momentum  of the ground state must be zero\cite{nafie1983,okuyama2009,bredtmann2015}; in other words, even if the hydrogen atom is moving, the electronic momentum is still calculated to be zero (which is incorrect).  
Moreover, in the context of energy, if we were to consider a system of $N$ isolated hydrogen atoms, the total difference in energy becomes $N$ times larger. Furthermore, if one were to bring those $N$ atoms closer together, so that they could interact and form hydrogen molecules or solids, there is no reason to assume that the problem of freezing nuclei (rather than choosing optimal coordinates) would go away\cite{bunker1977,moss1996,kutzelnigg2007,diniz2013,cederbaum2015,scherrer2017} -- in fact, one might expect that with more and more interacting nuclei, the problems of BO theory would only become more daunting and more difficult to solve.

\mycomment{Within the context of molecular quantum chemistry, it might appear that the problem above can be addressed by separating out the center of mass, in the spirit of Eq.~\ref{eq:Hcom} above\cite{bubin2013,Cederbaum2013,littlejohn2024}. 
Mathematically, for a system with an arbitrary number of electrons and nuclei, one can formally change coordinates and separate out the total center of mass by choosing Jacobi coordinates for the nuclei $\{\bm R_i\}$ and defining new  electrons coordinates $\bm r_a'$ relative to the nuclear center of mass ${\bm R}_{\rm com}$:
\begin{eqnarray} \label{eq:ncom}
    \hat {\bm R}_{\rm com} = \frac {\sum_A M_A {\hat {\bm R}_A}} {\sum_A M_A} , \quad \hat {\bm r}_a' = \hat {\bm r}_a - \hat {\bm R}_{\rm com}.
\end{eqnarray}
Note that here and throughout the paper, we will use Latin letters $A, B, C$ to label different atoms, $a,b,c$ to label different electrons and Greek letters $\alpha, \beta, \gamma$ for Cartesian coordinates.
By defining a coordinate transformation $(\bm R_A, \bm r_a) \to ({\bm R}_{\rm com}, \bm R_i, \bm r_a')$, the molecular Hamiltonian in Eq.~\ref{eq:Hgeneral} becomes:
\begin{eqnarray} \label{eq:Tgeneral}
    \hat H = \frac{\hat {\bm P}^2_{\rm com}} {2M_{\rm tot}} + \sum_i  \frac{\hat{\bm P}_i^2} {2 \mu_i} + \sum_a \frac {\hat {\bm p'}^2_{a}} {2 m_{\bm e}} + \sum_{a,b} \frac {\hat {\bm p'}_a \cdot \hat {\bm p'}_b} {2M_{\rm n}}
\end{eqnarray}
with the total nuclear mass $M_{\rm n} = \sum_A {M_A}$, the total molecular mass $M_{\rm tot} = M_{\rm n} + \sum_a {m_{\rm e}}$ and the reduced mass $\mu_i$ corresponding to the $i$-th internal nuclear Jacobi coordinate $\bm R_i$. 
The last term in Eq.~\ref{eq:Tgeneral} is often referred to as the mass polarization term \cite{davis1982} and a detailed definition of the internal coordinate $\bm R_i$ can be found in any standard textbooks\cite{cornille2003}. 

Now, it is crucial to emphasize that the transformations leading to Eq. \ref{eq:Tgeneral} above {\em do not solve} the problem of electronic momentum -- especially in the context of condensed phase chemistry with many molecules. In particular, consider a pair of H atoms separated far from each other, e.g. one on the earth and one on the moon.  The correct eigenspectrum is characterized by two good quantum numbers (one for each atomic level) plus the kinetic energy of each hydrogen atom:
\begin{eqnarray}\label{eq:Eexact2}
    E_{n_1,n_2}^{\rm exact}(\bm P_1,\bm P_2) = \frac{\bm P_1^2}{2(M + m_{\rm e})} + \frac{\bm P_2^2}{2(M + m_{\rm e})} -\frac{\mu e^4}{8 \epsilon_0^2 h^2} \frac{1}{n_1^2} -\frac{\mu e^4}{8 \epsilon_0^2 h^2} \frac{1}{n_2^2}.
\end{eqnarray}
Note that the  transformation in Eq.~\ref{eq:ncom} does not succeed at recovering this  (correct) electronic eigen-spectrum, but instead
introduces an irrelevant coordinate (the nuclear center of mass) which is somewhere in the exosphere. 
As a side note, in this article, we will ignore all relativistic and quantum electrodynamics effects and focus on effects beyond BO theory due to the finite nuclear mass.}

\section{Phase-Space Electronic Hamiltonian Theory}\label{sec:theory}
\subsection{Phase-Space Interpretation of the Hydrogen Atom}
\mycomment{New approaches that go beyond BO theory are clearly needed for accurate quantum chemistry\cite{pavanello2005,yonehara2012}. With this in mind, we have recently proposed a phase-space approach to treat the exact problem above. To motivate this approach, consider the hydrogen Hamiltonian 
in Eq.~\ref{eq:HHatom}. 
Let  $\hat U$ be the unitary operator that diagonalizes the electronic Hamiltonian 
in Eq.~\ref{eq:Helhydrogen}:
\begin{eqnarray} \label{eq:Vad}
\hat U\hat H^{\rm el}_{\rm Hydrogen} \hat U^\dagger &=& \hat V_{\rm ad}  \\
    {\hat H}_{\rm Hydrogen}^{\rm el} \ket{\psi_{n}}  &=&  E_n^{\rm BO} \ket{\psi_{n}}
\end{eqnarray}
As is well known\cite{mills2000}, the total Hamiltonian in this basis
\begin{eqnarray}
    \hH_{\rm ad} = \hat{U} \hH_{\rm hydrogen} \hat{U}^{\dagger}
\end{eqnarray}
has matrix elements that are given by:
\begin{eqnarray}
\label{eq:Hadmn}
    H_{{\rm ad}, mn} = \sum_{k} \frac {(\hat {\bm P}_{\rm n}\delta_{mk} - i\hbar {\bm d}_{mk})(\hat {\bm P}_{\rm n}\delta_{kn} - i\hbar {\bm d}_{kn})} {2M} + E_n^{\rm BO} \delta_{mn}
\end{eqnarray}
where the non-adiabatic coupling vector ${\bm d_{mk}} = \bra{\psi_m} \bm \nabla_{\bm R_{\rm n}} \ket{\psi_k}$.

Now, because of the translational invariance of the adiabatic basis $\{ \ket{\psi_{n}} \}$, i.e. 
\begin{eqnarray}
    \frac{\hbar}{i}\left( \frac{\partial}{\partial {\bm r}_e} +  \frac{\partial}{\partial {\bm R}_n} \right) \ket{\psi_n} = 
    \left( \hat {\bm P}_{\rm n} + \hat {\bm p}_{\rm e} \right) \ket{\psi_n} = 0, 
\end{eqnarray}
%
it follows that we can express Eq.~\ref{eq:Hadmn} as:
\begin{eqnarray} \label{eq:Hadmnp}
    H_{{\rm ad}, mn} = \sum_{k} \frac {(\hat {\bm P}_{\rm n}\delta_{mk} -  {\bm p}_{{\rm e}, mk})(\hat {\bm P}_{\rm n}\delta_{kn} - {\bm p}_{{\rm e},kn})} {2M} + E_n^{\rm BO} \delta_{mn}
\end{eqnarray}
with ${\bm p}_{{\rm e}, mk} = \bra{\psi_m} \hat {\bm p}_{\rm e} \ket{\psi_k}.$ 
Thus, we can write this operator in basis-free form as:
}
\begin{eqnarray} \label{eq:HadHatom}
    \hat H_{\rm ad}  = \frac{\hat{\bm P}_{\rm n}^2}{2M} + \hat{V}_{\rm ad}
    -\frac{  
    \hat{\bm P}_{\rm n} \cdot \hat{\bm p}_{\rm e}^{\rm ad} + \hat{\bm p}_{\rm e}^{\rm ad}  \cdot\hat{\bm P}_{\rm n} }{2M} + \frac{(\hat{\bm p}_{\rm e}^{\rm ad})^2}{2M}.  
\end{eqnarray}
where $\hat{\bm p}_{\rm e}^{\rm ad} = \hat U \hat {\bm p}_e \hat U^\dagger$.
\mycomment{Up to this point, there has been no approximation: Eq.~\ref{eq:HadHatom} is an exact representation of the hydrogen atom Hamiltonian that sets the stage for a phase-space approach.

If one now makes the ansatz that the nuclei are classical, one can construct a phase-space electronic Hamiltonian which is a function of the parameters (not operators) $\bm R_{\rm n}$ and $\bm P_{\rm n}$:
\begin{eqnarray}
    \hat{H}^{\rm ad}_{\rm PS}(\bR_{\rm n},\bP_{\rm _n})  = \frac{{\bm P}_{\rm n}^2}{2M} + \hat{V}_{\rm ad}(\bR_{\rm n})
    -\frac{  
    {\bm P}_{\rm n} \cdot \hat{\bm p}_{\rm e}^{\rm ad} + \hat{\bm p}_{\rm e}^{\rm ad}  \cdot{\bm P}_{\rm n} }{2M} + \frac{(\hat{\bm p}_{\rm e}^{\rm ad})^2}{2M}.  
\end{eqnarray}
At this point, and only for practical purposes, the most helpful next step is to transform the Hamiltonian back to the original (presumably diabatic) basis:
\begin{eqnarray}
    \hat{H}_{\rm PS} &=& \hat{U}^{\dagger} \hat{H}^{\rm ad} _{\rm PS} \hat{U} \\
    & = & \frac{{\bm P}_{\rm n}^2}{2M} +\hat{H}_{\rm el}
    -\frac{ {\bm P}_{\rm n} \cdot \hat{\bm p}_{\rm e}}{M} 
     + \frac{\hat{\bm p}_{\rm e}^2}{2 M}.  
\end{eqnarray}
A simple analysis shows that $\hat H_{\rm PS}$ represents a boost of the electron with nuclear momentum ${\bm P}_{\rm n}$, 
\begin{eqnarray} \label{eq:PSHatom}
    \hat{H}_{\rm PS} 
    & = & \frac{({\bm P}_{\rm n} - \hat{\bm p}_{\rm e})^2}{2M} +\hat{H}_{\rm el}, 
\end{eqnarray}
}
Henceforward, when we speak of phase-space electronic Hamiltonians, we will refer to Hamiltonians similar to Eq.~\ref{eq:PSHatom} above.  Note that the eigenvalues of Eq.~\ref{eq:PSHatom} are identical to the exact energies (up to a uniform constant shift) for a hydrogen atom 
\begin{eqnarray}
    E_{n}^{\rm PS}({\bm P}_{\rm n}) &=& \frac {\bm P_{\rm n}^2} {2(M + m_{\rm e})}  -\frac{\mu e^4}{8 \epsilon_0^2 h^2} \frac{1}{n^2}  
\end{eqnarray}
provided we equate ${\bm P}_{\rm n}$ (which captures the nuclear momentum) with $\bm P$ (which is formally the center of mass momentum).

The fact that we can generate the exact eigenvalues through Eq.~\ref{eq:PSHatom}  is very encouraging because the steps taken above offer an approach to generate phase-space Hamiltonians for systems with $N$ isolated hydrogen atoms. 
In such a case, all we must do is to partition the electronic momentum into different operators, one around each atom, so that one can effectively perform $N$ different local changes of coordinate around each atom $A$:
\begin{eqnarray}
    \hat{\bm p} \rightarrow \frac{1}{2} \sum_A \left( \hat{\theta}_A \hat{\bm p}+\hat{\bm p} \hat{\theta}_A \right)
\end{eqnarray}
Here, we define  $\hat{\theta}_A (\hat {\bm r})$ to be a partition of space:
\begin{eqnarray}
    \sum_A  \hat{\theta}_A(\hat {\bm r}) = 1.
\end{eqnarray}
Note that while there are many ways to choose the space partition function $\hat \theta_A$, a natural option is to choose  smooth functions that are localized on atoms (e.g., a combination of Gaussian functions).  

One can then make the ansatz that, for a general system with many electrons and nuclei, a phase-space electronic Hamiltonian will be of the form:
\begin{eqnarray}
    \hat H_{\rm PS} &=& \sum_A \frac{{\bm P_A}^2}{2M_A}    
    - \frac 1 2 \sum_A \frac {\bm P_A}  {M_A}   
      \cdot   \left( \hat{\theta}_A \hat{\bm p} + \hat{\bm p} \hat{\theta}_A  
  \right) \nonumber
       \\
     &+& \frac 1 4 \sum_A \frac{\left(\hat{\theta}_A \hat{\bm p} + \hat{\bm p} \hat{\theta}_A \right)^2}{2 M_A } +\hat{H}_{\rm el}({\bm R}) 
\end{eqnarray} 
or equivalently
\begin{eqnarray} \label{eq:HPS'}
   \hat H_{\rm PS} &=&    \sum_A \frac{{\left(\bm P_A - i\hbar \hat {\bm \Gamma}_A'\right)}^2}{2M_A} +\hat{H}_{\rm el}({\bm R})
\end{eqnarray}
with
\begin{eqnarray} \label{eq:gamma1}
    \hat{\bm \Gamma}_A' = \frac 1 {2i\hbar} \left( \hat{\theta}_A \hat{\bm p} + \hat{\bm p}   \hat{\theta}_A\right).
\end{eqnarray}
To our knowledge, Eqs.~\ref{eq:HPS'} and \ref{eq:gamma1} offer the simplest form for a many-electron/many-nuclei phase-space electronic Hamiltonian with clear physical meaning. 
 
\subsection{Phase-Space Electronic Hamiltonians and a Simple Form of $\hat {\bm \Gamma}_A$}

Eq.~\ref{eq:HPS'} above was guessed on the basis of the hydrogen atom.  To better  understand where such an equation comes from, and to better improve upon it, note that, if one diagonalizes $\hat H_{\rm el}\left(\bm R\right)$ in Eq.~\ref{eq:HPS'},  the {\em exact} Hamiltonian in the basis of electronic eigenstates $\{\ket{\psi_j} \} $ is:
\begin{eqnarray} \label{eq:Hadiabatic}
    \hat{H}_{\rm ad} = \sum_A \frac{ {\left(\hat{\bm P}_A - i\hbar \hat{\bm d}_A\right)}^2}{2M_A} +\hat{V}_{\rm ad} (\hat {\bm R})
\end{eqnarray}
where \mycomment{$\hat {\bm d}_A = \sum_{jk} {\bm d}_{jk}^A \ket{\psi_j} \bra{\psi_k}$ is the derivative coupling operator and $\hat V_{\rm ad}$ is the diagonal matrix of eigenvalues from Eq.~\ref{eq:Vad}.
Thus, a phase-space approach is an attempt to capture some of the beyond BO physics present in Eq.~\ref{eq:Hadiabatic} by finding an operator $\hat \bGamma$ that mimics the derivative coupling $\hat \bd$.

Now, formally, the derivative coupling operator $\hat {\bm d}_A$ is not an operator so much as a (expensive) matrix of elements in the adiabatic basis. That being said, it is not difficult to show that ${\bm d}^A_{jk}$ satisfies four properties:
\begin{eqnarray}
    -i\hbar\sum_{A} {\bm d}^A_{jk} + \bra{\psi_j}\hbm{p} \ket{\psi_k} &=& 0,\label{eq:dconstrain1}  \\
    \sum_A \bm \nabla_{A} {\bm d}^B_{jk}  &=& 0 
    \label{eq:dconstrain2}\\
    -i\hbar\sum_{A}{\bm R}_{A} \times {\bm d}^A_{jk} + \bra{\psi_j} \hbm{l} + \hbm{s} \ket{\psi_k} &=& 0,\label{eq:dconstrain3}\\
     -\sum_A {\bm R}_A \times \bm \nabla_{A} {d}^{B\beta}_{jk} &
     =&  \sum_{\gamma} \epsilon_{\alpha \beta \gamma} d^{B \gamma}_{jk},\label{eq:dconstrain4} 
\end{eqnarray}
where $\hat{\bm l}$ and $\hat{\bm s}$ are the electronic orbital and spin angular momentum operators.
In words, Eqs. \ref{eq:dconstrain1} and \ref{eq:dconstrain3} reflect the fact that the system is unchanged under total translation and rotation of the entire system (of electrons and nuclei).  Eqs. \ref{eq:dconstrain2} and \ref{eq:dconstrain4} reflect that the derivative coupling vectors themselves transform correctly under translation and rotation. 

Now, in developing a meaningful phase-space approach of the form 
\begin{eqnarray} \label{eq:HPS}
   \hat H_{\rm PS} &=&    \sum_A \frac{{\left(\bm P_A - i\hbar \hat {\bm \Gamma}_A\right)}^2}{2M_A} +\hat{H}_{\rm el}({\bm R}), 
\end{eqnarray}
one criterion for constructing a meaningful $\hat \bGamma$ operator is that the operator should satisfy the parallel constraints:

\begin{eqnarray}
    -i\hbar\sum_{A}\hbm{\Gamma}_{A} + \hbm{p} &=& 0,\label{eq:constrain1}  \\
    \Big[-i\hbar\sum_{B}\pp{}{\bm{R}_B} + \hbm{p}, \hbm{\Gamma}_A\Big] &=& 0,\label{eq:constrain2}\\
    -i\hbar\sum_{A}{\bm R}_{A} \times \hat{\bm \Gamma}_{A} + \hbm{l} + \hbm{s} &=& 0,\label{eq:constrain3}\\
     \Big[-i\hbar\sum_{B}\left(\bm{R}_B \times\pp{}{\bm{R}_B}\right)_{\gamma} + \hat{l}_{\gamma} + \hat{s}_{\gamma}, \hat{\Gamma}_{A \delta}\Big] &
     =& i\hbar \sum_{\alpha} \epsilon_{\alpha \gamma \delta} \hat{\Gamma}_{A \alpha},\label{eq:constrain4} 
\end{eqnarray}
Moreover, one can show that dynamics along an eigenstate of a Hamiltonian of the form in Eq.~\ref{eq:HPS} (where $\hat{\bm \Gamma}$ satisfies Eqs.~\ref{eq:constrain1}-\ref{eq:constrain4})  will conserve linear and angular momentum\cite{qiu2024,tao2024,tao2024:bf}.
}

Alas, $\hat{\bm \Gamma'}$ in Eq.~\ref{eq:gamma1} satisfies only Eqs.~\ref{eq:constrain1} and \ref{eq:constrain2}, but not Eqs.~\ref{eq:constrain3} and \ref{eq:constrain4}.
The implication of this fact is that,
if one diagonalizes Eq.~\ref{eq:HPS'}, then classical dynamics propagated along the resulting energy surfaces will  conserve linear momentum\cite{tao2024:bf} but not angular momentum.
Nevertheless, we have argued that one can use $\hat \bGamma'$ in Eq.~\ref{eq:gamma1} as a starting point, and
 build a proper $\hat{\bm \Gamma}$ as the sum of two contributions:
\begin{equation}
    \hat{\bm \Gamma} =  \hat{\bm \Gamma}' +  \hat{\bm \Gamma}''
\end{equation}
with final form of the phase-space Hamiltonian being \cite{qiu2024}:
\begin{eqnarray}
\label{eq:hps:general}
    \hat{H}_{\rm PS} &=&   \sum_A \frac{ {\left(\bm P_A - i\hbar  \hat {\bm \Gamma}_A' - i\hbar \hat {\bm \Gamma}_A'' \right)}^2}{2M_A} +\hat{H}_{\rm el}({\bm R})
\end{eqnarray}

All that remains is to build the $\hat {\bm \Gamma}'$ and $\hat {\bm \Gamma}''$  operators, which can be thought of as the local translational and rotational components of the derivative couplings\cite{thorson1978,illescas1998,fatehi2012,athavale2023,tao2024:bf}. 
Let us now describe how to construct these objects explicitly.

\subsubsection{Form of $\hat {\bm \Gamma}_A'$}
For the construction of $\hat{\bm \Gamma'}$  in Eq. \ref{eq:gamma1}, it remains only to construct the partition of unity function $\hat \theta$.
We have chosen to define $\hat \theta_A$ as combinations of Gaussian functions localized on different atoms: 
\begin{eqnarray}
\label{eq:theta}
    \hat \theta_A(\hat{\bm r}) = \frac{M_Ae^{-|\hat{\bm r}-\bm{R}_A|^2/\sigma_A^2}}{\sum_B M_Be^{-|\hat{\bm r}-\bm{R}_B|^2/\sigma_B^2}}.
\end{eqnarray}
In Eq.~\ref{eq:theta}, we utilize  Gaussian functions with width $\sigma_A$, centered around $\bm R_A$ and weighted by $M_A$.  
The physical meaning of this spatial partition function $\hat \theta_A$ is then clear; $\hat \theta_A$ measures the extent to which the electrons are influenced by nucleus $A$, i.e., electrons are more strongly affected by nucleus $A$'s motion when they are closer to $A$ or $A$ is heavier.  
Note that the parameter $\sigma_A$ determines the locality of this measure. In practical {\em ab initio} electronic structure calculations, as discussed in Ref.~\citenum{tao2024:bf}, an optimal choice for $\sigma_A$ can be $\sigma_A = 0.2 r_{\rm vdW}^A$, where $r_{\rm vdW}^A$ is the van der Waals radius of atom $A$.

\subsubsection{Form of $\hat {\bm \Gamma}_A''$}
For the rotational component $\hat {\bm \Gamma}_A''$, we define the $\hat {\bm \Gamma}_A''$ to be the basis-free electronic rotation factor: 
\begin{eqnarray}
    \hat{\bm \Gamma}_A'' &=& \sum_B\zeta_{AB}\left(\bm{R}_A -\bm{R}^0_{B}\right)\times \left(\bm{K}_B^{-1}\hat{\bm J}_B\right),\label{eq:erf_final}
\end{eqnarray}
where 
\begin{eqnarray}
    \hat{\bm J}_B &=& \frac{1}{2i\hbar}\left((\hat{\bm r}-\bm{R}_B)\times \hat \theta_B(\hat{\bm r} )\hat{\bm p} +(\hat{\bm r} -\bm{R}_B)\times\hat{\bm p} \hat \theta_B(\hat{\bm r})+2\hat{\bm s} \hat{\theta}_B\right), \\
    \bm{R}_{B}^0 &=& \frac{\sum_A\zeta_{AB}\bm{R}_A}{\sum_A\zeta_{AB}},\\
    \label{eq:KB}
    \bm{K}_B &=& \sum_A\zeta_{AB}\left(\bm{R}_A\bm{R}_A^\top-\bm{R}_B^0\bm{R}_B^{0\top}-(\bm{R}_A^\top\bm{R}_A-\bm{R}_B^{0\top}\bm{R}_B^0)\mathcal{I}_3\right),
\end{eqnarray}    
and the symmetric localization function $\zeta_{AB}$
\begin{eqnarray}
\label{eq:zeta}
    \zeta_{AB} &=e^{-\frac {|\bm{R}_A-\bm{R}_B|^2} {2(\sigma_A + \sigma_B)^2}}.
\end{eqnarray} 

It is straightforward to verify that the $\hat {\bm \Gamma}$ term defined above satisfies the following four constraints in Eqs.~\ref{eq:constrain1}-\ref{eq:constrain4}\cite{tao2024:bf}.
Altogether, Eqs.~\ref{eq:hps:general}-\ref{eq:zeta} constitute a complete, self-contained, practical phase-space Hamiltonian $\hat H_{\rm PS}$.

\subsection{Phase-Space Electronic Hamiltonian and Its Connection to Berry's Superadiabatic Formalism} 
\mycomment{
At this point, we have sought to justify the use of an  electronic Hamiltonian by noting that the approach includes some non-Born Oppenheimer effects by approximating the derivative coupling ($\hat {\bm d}$) with an operator ($\hat{\bm \Gamma}$).  That being said, it is important to note that a similar conclusion can also be drawn by considering Berry's superadiabatic formalism--which is not very well known in chemistry and  which we now briefly review. 

Consider a Hamiltonian $\hat H(\lambda(t))$ with a slow varying time-dependent parameter $\lambda$. Let us diagonalize the Hamiltonian to obtain the zeroth-order adiabatic basis $\{ \ket{\psi^0_j} \}$,  
\begin{eqnarray}
    \hat H \ket{\psi^0_j} = E^0_j \ket{\psi^0_j}.
\end{eqnarray}
and expand the system wavefunction in this adiabatic basis,
\begin{eqnarray}
    \ket{\Psi} = \sum_j c_j^0 \ket{\psi^0_j}. 
\end{eqnarray}
In this basis, the time-dependent Schrodinger equation can  be expressed as
\begin{eqnarray} \label{eq:tdse0}
    \dot {c}_j^0  = -\frac i \hbar E_j^0 c_j^0 + \sum_k \dot \lambda \bra{\psi_j^0} \nabla_\lambda \ket{\psi_k^0} c_k^0.
\end{eqnarray}
Thus, an effective first-order Hamiltonian can be identified from Eq.~\ref{eq:tdse0} as:
\begin{eqnarray}
    H^1_{jk} =  E_j^0 - i\hbar \dot \lambda \bra{\psi_j^0} \nabla_\lambda \ket{\psi_k^0},
\end{eqnarray}
As noted by Ref.~\citenum{berry1990}, one can diagonalize this Hamiltonian to generate a first-order super-adiabatic basis
\begin{eqnarray}
    \hat H^1 \ket{\psi_j^1} = E_j^1\ket{\psi_j^1}. 
\end{eqnarray}
Moreover, as the first-order superadiabatic basis remains time-dependent, this diagonalization process can be applied iteratively to construct higher and higher-order superadiabatic bases.

Now, the hope is that by performing such iterative diagonalizations, one can arrive at a superadiabatic basis, whereby non-adiabatic effects can be effectively suppressed (even if it will not completely vanish), making the propagation of adiabatic dynamics along superadiabatic surface more accurate.   Moreover, in the context of semiclassical molecular dynamics, in a landmark paper\cite{shenvi2009}, Shenvi showed that, to first order only, this approach can be used to generate phase-space electronic Hamiltonians of the form:
\begin{eqnarray}  \label{eq:HPSShenvi}
    H_{{\rm PS}, mn}^{\rm Shenvi} (\bm R, \bm P) =  
    \sum_{k} \frac {({\bm P}_A \delta_{mk}  - i\hbar{\bm d}^A_{mk})({\bm P}_A\delta_{kn} - i\hbar{\bm d}^A_{kn})} {2M_A} + E_n^{\rm BO}(\bR) \delta_{mn}
\end{eqnarray}
Shenvi posited that surface hopping dynamics along these surfaces can be used for so-called phase-space surface hopping (PSSH) dynamics, and Ref.\citenum{shenvi2009} demonstrates that for some particular cases, PSSH can vastly outperform more standard fewest switches surface hopping dynamics (FSSH)\cite{tully1990}.

At this point, however, it is worth noting that there are several problems with Eq.~\ref{eq:HPSShenvi}. In particular, this Hamiltonian becomes problematic at the vicinity of conical intersections, where $\bm d_{jk}$ diverges. This Hamiltonian is also  computationally expensive,  as calculating the forces on the superadiabatic surfaces would appear to  require evaluating second-order derivatives (the derivatives of derivative couplings). Finally, treating degenerate adiabatic states (with potentially arbitrary gauge choices) becomes unreliable; in such a case, the derivative couplings are not even well-defined. 
For all of the reasons above, very little work has been achieved using an electronic phase-space Hamiltonian of the form in Eq.~\ref{eq:HPSShenvi} beyond model problems\cite{gherib2016}. Nevertheless, as we have argued above, if one can approximate $\hat{\bm d}$ by constructing a meaningful (and tractable) $\hat \bGamma$  operator, we believe there is a strong future for electronic phase-space Hamiltonian of the form in Eq.~\ref{eq:hps:general}.
}

\subsection{Our Goal In this Article}

This concludes our heuristic review of phase-space electronic Hamiltonian theory. 
We have sought to justify a phase-space approach  by appealing to the readers' understanding of the importance of a change of coordinates when simulating a simple hydrogen atom, but have also demonstrated that the approach is reasonable by mapping the approximate $\hat {\bm \Gamma}$ factors to the derivative coupling. 
By this point in time, there is already ample evidence in the literature that this phase-space approach can recover electronic momenta\cite{tao2024}, electronic current densities\cite{tao2024:bf} and even vibrational circular dichroism (VCD) signals\cite{duston2024}.  
In all cases, we have found that a phase-space electronic Hamiltonian is simply more accurate, and offers more and richer set of information than a BO Hamiltonian.

That being said, however, one could argue that we still do not have a smoking gun that clearly establishes the supremacy of the phase-space approach. Furthermore, one might also note that the phase-space approach developed in Sec.~\ref{sec:theory} does require some hand-waving, and for complete confidence, one would like to understand how to derive and work with such a Hamiltonian in the context of an exact calculation. These concerns motivate the present article. In what follows, we will  first establish how a phase-space electronic Hamiltonian can be rigorously defined through a Wigner transformation, and then, using the inverse Weyl transformation, we will show how to extract quantized vibrational energies from a phase-space approach. In particular, we will show that vibrational energies extracted by diagonalizing the nuclear Hamiltonian from a phase-space electronic Hamiltonian universally outperforms (i.e. gives more accurate vibrational energies) than  diagonalizing the corresponding nuclear Hamiltonian from a Born-Oppenheimer electronic Hamiltonian. This finding would appear to be the smoking gun that makes clear that, in the future,  quantum chemistry packages will need to be prepared to  avoid the Born-Oppenheimer Hamiltonian.

\section{Born-Oppenheimer  vibrational energies}
Before demonstrating how to extract vibrational energies from phase-space theory, it will be helpful to review briefly the easiest way to derive BO theory and BO vibrational energies from a formal point of view for a system of many electrons and nuclei (and thus expand on the simple approach for the hydrogen atom in Sec.~\ref{sec:intro} above).


The usual BO {\em framework} proceeds by initially defining a unitary operator $\hat U$ 
that diagonalizes $\hat H_{\rm el}$. One then transforms to the new adiabatic basis,
\begin{eqnarray} \label{eq:Had}
    \hat H_{\rm ad} &=& \hat U \hat H \hat U^{\dagger} =  \frac {(\hat {\bm P}- i\hbar \hat {\bm d})^2} {2M} + \hat V_{\rm ad}(\hat {\bm R}).
\end{eqnarray} 
The BO {\em approximation} is then that one can neglect the $\hat {\bm P} \cdot \hat{\bm d}$ and $\hat {\bm d} \cdot \hat{\bm d}$ terms in Eq.~\ref{eq:Had}, and diagonalize:
\begin{eqnarray}
\label{eq:HBO:easy}
    \hH_{\rm BO}  = \frac {\hat {\bm P}^2} {2M} + \hat V_{\rm ad}(\hat{\bm R}).
\end{eqnarray}
Note that, because $\hat V_{\rm ad} = {\rm diag}(E_0(\hat \bR), \cdots, E_n(\hat \bR) )$, 
the BO assumption amounts to many different diagonalizations of $\frac {\hat {\bm P}^2} {2M} + E_n(\bR)$.  Note also that,  while $\hat H_{\rm ad}$ is a unitary transform of $\hat H$, $\hat H_{\rm BO}$ is clearly not such a transform.

Everything above is well known and standard. Also well known, but slightly less standard, is that the BO approximation can also be derived through partial Wigner transforms\cite{wigner1932, balazs1984, kapral1999}. 
As a reminder, the partial Wigner transform ($\mathcal{W}$) for any operator $\hat{O}$ is given by:
\begin{eqnarray} \label{eq:wigner}
    \hat O_W(\bR,\bP) &=& \mathcal{W}\left[ \hat{O} \right]\\
    & \equiv & \int d\bm R' \bra{\bm R + \frac {\bm R'} 2 } \hat O \ket{\bm R - \frac {\bm R'} 2 } e^{-\frac i \hbar \bm R' \cdot \bm P},
\end{eqnarray}
where $\hat O_W(\bR,\bP)$ is a matrix of electronic operators that depends parametrically on $\bR$ and $\bP$. This transformation is invertible by a Weyl transform, and one can show that:
\begin{eqnarray}
     \hat{O}  &=& \mathcal{W}^{-1} \left[ \hat{O}_W \right],
\end{eqnarray}
with matrix elements
\begin{eqnarray}\label{eq:weyl}
    \bra{\bm R} \hat{O} \ket{\bm R'}  
    & = & \int \frac {d\bm P} {2\pi\hbar} e^{\frac i\hbar \bm P \cdot (\bm R - \bm R')} \hat O_W\left(\frac {\bm R + \bm R'} 2, \bm P\right) .
\end{eqnarray}

Now, to formally derive the BO approximation, we build and diagonalize the partial Wigner transform of $\hH$:
\begin{eqnarray} \label{eq:HW}
    \hH_W  &=&  \frac {\bm P^2} {2M} + \hH_{\rm el}(\bm R) =  \hat{U}_W^{\dagger}\hat \Lambda_W \hat{U}_W.
\end{eqnarray}
\mycomment{Note that, on the right hand side of Eq.~\ref{eq:HW}, $\hat{U}_W$  is an electronic unitary operator, parametrized by nuclear position  $\bR$ and we  perform electronic matrix multiplication. }
The BO Hamiltonian in Eq.~\ref{eq:HBO:easy} is then exactly equivalent to the Weyl transform of $\Lambda_W$:
\begin{eqnarray}
\label{eq:HBO:W}
    \hat H_{\rm BO} = \mathcal{W}^{-1} \left[\hat \Lambda_W \right] = \mathcal{W}^{-1} \left[ \hat U_W \hH_W \hat U_W^{\dagger} \right].
\end{eqnarray}
Note that the BO Hamiltonian is (obviously) not a unitary transform of the exact Hamiltonian. If one were to apply a partial Wigner transformation to the adiabatic Hamiltonian in Eq.~\ref{eq:Had}, one must remember that the Wigner transform of a product of operators is not the product of Wigner transforms but rather involves a Moyal product between them:   
\mycomment{
\begin{eqnarray}
    \left( \hat A \hat B \right)_W = A_W * B_W.
\end{eqnarray}
Here,  the Moyal product between two phase-space functions $A_W(\bm R, \bm P)$ and $B_W(\bm R, \bm P)$ is defined as 
\begin{eqnarray}
    A_W(\bm R, \bm P) * B_W(\bm R, \bm P) = A_W(\bm R, \bm P) e^{-\frac {i \hbar} 2 \hat \Lambda} B_W(\bm R, \bm P)
\end{eqnarray}
where the Possion bracket operator $\hat \Lambda = \cev{\bm \nabla}_{\bm P} \cdot \vec {\bm \nabla}_{\bm R} -  \cev{\bm \nabla}_{\bm R} \cdot \vec {\bm \nabla}_{\bm P}$.
}
In such a case, it follows that the exact Hamiltonian can be written as: 
\begin{eqnarray}\label{eq:Had:W}
    \hat H_{\rm ad} &=& \mathcal{W}^{-1} \left[\hat U_W * \hH_W * \hat U_W^{\dagger} \right] = \mathcal{W}^{-1} \left[ \left(\hat U\hH \hat U^{\dagger}\right)_W\right].
\end{eqnarray}
Observe that the difference between Eq.~\ref{eq:HBO:W} and Eq.~\ref{eq:Had:W} is that normal product multiplication is replaced by Moyal  products.

In any event, it is quite common today to extract vibrational energies from a BO Hamiltonian. 
To do so, for the ground state, one simply diagonalizes the nuclear Hamiltonian for the electronic ground state $\bra{\psi_0} \hat{H}_{\rm BO} \ket{\psi_0}$ (which is equivalent to diagonalizing $\frac {\hat {\bm P}^2} {2M} + E_0(\hat \bR)$). Alternatively,  within a harmonic approximation, one can can simply expand $E_0$ (or equivalently $ \bra{\psi_0} \hat \Lambda_W \ket{\psi_0} $) to second order in $\bm R$ and apply the standard harmonic oscillator result\cite{neugebauer2002}.


\section{Vibrational Energies from a Phase-Space Electronic Hamiltonian}
According to the theory presented in Sec.~\ref{sec:theory}, there are better means to approximate the full Hamiltonian than $\hat H_{\rm BO}$.
In generating such approximations, and in light of Eq.~\ref{eq:hps:general}, it would appear that the optimal approach is to parameterize the approximate Hamiltonian by both nuclear position and momentum. Rigorously, such a parameterization can only occur after a Wigner transformation.  
Thus, the phase-space approach is to construct an electronic Hamiltonian of the form:
\begin{eqnarray} \label{eq:hps:w}
    \hH_W^{\rm PS} &=& \hH_W -i\hbar \hat{\bm \Gamma}_W \cdot \frac{\bm P}{M} - \hbar^2\frac{\hat{\bm \Gamma}_W^2}{2M},
\end{eqnarray}
and then diagonalize it:
\begin{eqnarray}
    \hH^{\rm PS}_W &=& L_W^{\dagger}  \hat {\Lambda}^{\rm PS}_W L_W.
    \label{eq:diagps:lam}
\end{eqnarray}



\mycomment{As above for the BO case, on the right hand side of Eq.~\ref{eq:diagps:lam}, $\hat{L}_W$  is an electronic unitary operator -- but now parametrized by both nuclear position  $\bR$ and nuclear momentum $\bP$ -- and  we  perform electronic matrix multiplication in Eq.~\ref{eq:diagps:lam}. }
The final phase-space electronic Hamiltonian is  then constructed (as for BO) by a Weyl transformation:
\begin{eqnarray}
\label{eq:HPS:W}
\hH_{\rm PS} &=& \mathcal{W}^{-1} \left[\Lambda_W^{\rm PS}\right]  = \mathcal{W}^{-1} \left[ L_W H_W^{\rm PS} L_W^{\dagger}  \right]
\end{eqnarray}
Eq.~\ref{eq:HPS:W} should be compared with Eq.~\ref{eq:HBO:W}.  Note that, just like $\hH_{\rm BO}$, $\hH_{\rm PS}$ is {\em not} a unitary transform of $\hH$.
Also, just as for $H_{\rm BO}$, vibrational energies can  be formally extracted by diagonalizing the nuclear Hamiltonian $\bra{\psi_0^{\rm PS}} \hat{H}_{\rm PS} \ket{\psi_0^{\rm PS}}$; however,  unlike the BO case, constructing $\hat{H}_{\rm PS}$ {\bf requires} a Weyl transform in Eq.~\ref{eq:HPS:W}. Put differently, while working with the phase-space electronic Hamiltonian is obvious for semiclassical simulations, for quantum nuclear (i.e. vibrational) problems, where $\bP$ is an operator (not a paramter), the phase-space electronic Hamiltonian has no simple analogue to Eq.~\ref{eq:HBO:easy}
(but rather requires a Weyl transformation).

As discussed above, previous work has demonstrated that a phase-space approach has many advantages over a BO Hamiltonian -- one recovers the exact eigenvalues of a hydrogen atom, one recovers nonzero electronic momenta and electronic current density,  and one can also match experimental VCD spectroscopy. 
In what follows, we will now show that, using the Hamiltonian in Eq.~\ref{eq:HPS:W}, we also recover improved vibrational energies (often much improved).

Before concluding this section, one item must be emphasized. For the  reader who is scared of Wigner transforms, note that a harmonic phase-space vibrational energy (which can be calculated by a quadratic expansion and entirely without a Wigner transformation) always outperforms a harmonic BO calculation for vibrational energies.  As such, there is not only a conceptual but also a very practical reason to employ a phase-space Hamiltonian in the immediate future.

\section{Demonstration through a model problem}
To demonstrate the power of a phase-space approach for calculating vibrational spectra, consider a classic model Hamiltonian that describes a one-dimensional asymmetric hydrogen bond system $\rm O-H\cdots O$ from Refs.\citenum{marinica2006,scherrer2017}. 
In this coarse-grained model, the light particle represents a hydrogen atom with mass $m$, and the heavy particles are oxygen atoms with mass $M$ -- though we will also refer to the particle as an electron and the heavy particle as a nucleus, given the importance of BO theory to electronic structure calculations.

\subsection{Model Hamiltonian}
The Hamiltonian of the three-particle one-dimension hydrogen bond model can be written as:
\begin{eqnarray} \label{eq:fullH}
   \hat H =  \frac {\hat P_1^2} {2M} + \frac {\hat P_2^2} {2M} + \frac {\hat p^2} {2m} +  \hat V(\hat R_1, \hat R_2, \hat r_{\rm e}),
\end{eqnarray} 
where we assume the mass of two heavy particles are the same $M_1 = M_2 = M$. 
We further reduce the complexity of this Hamiltonian by going to the molecular (nuclei + electron) center of mass frame. 
Namely, we define the molecular center of mass (MCM)
\begin{eqnarray} 
    \hat R_{\rm MCM} =  \frac {M\hat R_1 + M\hat R_2 + m\hat r_{\rm e}} {2M + m}, 
\end{eqnarray} 
We  transform the coordinates from $(\hat R_1, \hat R_2, \hat r_{\rm e})$ to MCM coordinates $(\hat R_{\rm MCM}, \hat R, \hat r)$,
where $\hat R$ is the distance between the two oxygen atoms and the light degrees of freedom $\hat r$ indicates the distance between the H atom to the nuclear center of mass.
\begin{equation}
    \hat R =  \hat R_1 - \hat R_2, \quad 
  \hat r =  \hat r_{\rm e} -  \frac{\hat R_1 + \hat R_2}{2}.  
\end{equation}
The Hamiltonian therefore becomes effectively a two dimensional model if we get rid of the molecular center of mass motion:
\begin{eqnarray} \label{eq:modelH}
    \hat H =  \frac {\hat P_{\rm MCM}^2} {2(2M + m)} + \frac {\hat P^2} {2\mu}  + \frac {{\hat {p}}^2} {2m} +  \frac {{\hat {p}}^2} {2(M+M)}
    + \hat V(\hat R, \hat r), 
\end{eqnarray} 
where $\mu = \frac M 2$ is the nuclear reduced mass and the second to last term is the mass polarization term, 
and the momentum operators in the new coordinate system are defined as:
\begin{eqnarray} 
    \hat P_1 = \frac {M \hat P_{\rm MCM}} {2M+ m} + \hat P - \frac {\hat p} {2}, \quad  \hat P_2 = \frac {M \hat P_{\rm MCM}} {2M+ m} - \hat P - \frac {\hat p} {2}, \quad \hat p_{\rm e} = \hat  p + \frac {m \hat P_{\rm MCM}} {2M + m}.
\end{eqnarray} 

In the MCM coordinates, the potential energy term does not depends on the $R_{\rm MCM}$ and can be expressed as:
\begin{eqnarray} \label{eq:modelV}
\hat V(\hat R, \hat r) &=& D\left(e^{-2a\left(\frac {\hat R} 2 + \hat r - d\right)} -2e^{-a\left(\frac {\hat R } 2+ \hat r - d\right)} + 1 \right) \nonumber \\
&+& Dc^2\left(e^{-\frac {2a} c\left(\frac {\hat R} 2 - \hat r - d\right)} -2e^{-\frac a c\left(\frac {\hat R} 2 - \hat r - d\right)}\right) + Ae^{-B\hat R} - \frac C {\hat R^6}. 
\end{eqnarray}
where the parameters used in the paper are given in Table.~\ref{tab:parameters}.
\begin{table}[h]
    \centering
    \begin{tabular}{|c|c|c|c|c|c|c|c|}
        \hline
        $D$ & $d$ & $a$ & $c$ & $A$ & $B$ & $C$ \\
        \hline
        $60~\rm kcal/mol$ & $0.95~\textup{\AA}$ & $2.52~\textup{\AA}^{-1}$ & $0.707$ & $2.32 \times 10^5~\rm kcal/mol$ & $3.15~\textup{\AA}^{-1}$ & $2.31 \times 10^4~\rm kcal/mol/\textup{\AA}^{6}$ \\
        \hline
    \end{tabular}
    \caption{Model parameters from Ref.~\citenum{marinica2006}.}
    \label{tab:parameters}
\end{table}

The potential from Refs.\citenum{marinica2006,scherrer2017}  is plotted in Fig.~\ref{fig:PES}(a) as a function of the heavy and light degrees of freedom $R$ and $r$. The intuition behind this model Hamiltonian is that the hydrogen will stay at the center when the two oxygen atoms are close, and it will bonded to one oxygen when the oxygen atoms are far away due to the hydrogen bonding effects.  

\subsection{Methods}
For this model Hamiltonian, we have calculated vibrational data using three different approaches:

\subsubsection{Exact calculations} 
For the exact results, we diagonalize $\hH$ over a joint $(N_r \times N_R) \times(N_r\times N_R) $ grid in the nuclear coordinate $R$ and the electronic coordinate $r$ , where we choose $N_r = 400$ grids points for the electronic coordinate $r \in \left[-2, 2\right] \textup{\AA}$ and $N_R = 100$ grid points for nuclear coordinate $R \in \left[2, 4\right] \textup{\AA}$.
Here and below (for all methods), we constructed the electronic momentum and electronic kinetic energy operators by the finite-difference stencil technique, while  the nuclear kinetic operator is built with a Fourier transform using the conjugate momentum grid points.

\subsubsection{BO calculations}
For the BO results,  we use the same grids as above the for the exact calculations (with $N_r = 400$ grids points for the electronic coordinate $r \in \left[-2, 2\right] \textup{\AA}$ and $N_R = 100$ grid points for nuclear coordinate $R \in \left[2, 4\right] \textup{\AA}$).  As is standard, we begin by looping over the grid of nuclear positions $R_i$;  thereafter we use the grid of electronic positions to diagonalize the electronic Hamiltonian associated for each nuclear grid point $\hH_{\rm el}(R_i) \ket{\psi_0} = E_0(R_i) \ket{\psi_0}$. Finally, we  diagonalize the nuclear Hamiltonian $\frac {\hP^2} {2M} + E_0(\hat R)$ over the nuclear grid to generate vibrational energies.


\subsubsection{Phase-space Calculations} \label{sec:flowchart}

To generate the phase-space results, we first construct the phase-space electronic Hamiltonian for the three-particle model system in the MCM coordinates. 
Since the model Hamiltonian only has one-spatial dimension, we do not need to consider the rotational part of $\hat {\bm \Gamma}$ operator ($\hat{\bm \Gamma}''$). 
In the original coordinates, the phase-space Hamiltonian can be expressed as: 
\begin{eqnarray}
\label{eq:fullHPS}
   \hat H_{\rm PS} (R_1,R_2, P_1, P_2) = \frac {(P_1 - i\hbar\hat \Gamma_1)^2} {2M} +  \frac {(P_2 - i\hbar\hat \Gamma_2)^2} {2M}+ \frac {\hat p^2} {2m} +  \hat V(R_1, R_2, \hat r_{\rm e}).
\end{eqnarray}
At this point, it will be convenient to define the nuclear center of mass (NCM) coordinates: 
\begin{eqnarray} \label{eq:transform}
    \hat R_{\rm NCM} = \frac {\hat  R_1 + \hat R_2} {2}.
\end{eqnarray}  
If we change coordinates from $(R_1,R_2)$ to $(R,R_{NCM})$, the result is:
\begin{eqnarray}
\label{eq:complicated}
   \hat H_{\rm PS} (R,P,P_{NCM}) =
   \frac{(P_{NCM}+ \hat{p})^2}{2(M+M)} +
   \frac {\left( P- i\hbar \frac{(\hat \Gamma_1 - \hat \Gamma_2)}{2} \right)^2} {2\mu} + \frac {\hat p^2} {2m} +  \hat V(R , \hat r).
\end{eqnarray}
We can identify the $\hat{\Gamma}$ operator for the relative $R$ coordinate, 
\begin{eqnarray} 
      \hat \Gamma  = \frac {\hat \Gamma_1 - \hat \Gamma_2} 2.
\end{eqnarray}  
At this point, we ignore the MCM motion encapsulated in the first term in Eq.~\ref{eq:complicated}, so that the final phase-space electronic Hamiltonian becomes
\begin{eqnarray}
\label{eq:HPS:model}
   \hat H_{\rm PS} (R,P) = \frac {(P- i\hbar \hat \Gamma)^2} {2\mu} + \frac {\hat p^2} {2m} +  \hat V(R , \hat r), 
\end{eqnarray}
where
\begin{eqnarray}
    \hat \Gamma_1 =  \frac{1}{2i\hbar}\left( \hat{\theta}_1\hat{ p}  + \hat{p}  \hat{\theta}_1\right), \quad \hat \Gamma_2 =  \frac{1}{2i\hbar}\left( \hat{\theta}_2\hat{p}  + \hat{p}  \hat{\theta}_2\right)
\end{eqnarray}
with 
\begin{eqnarray} \label{eq:theta:used}
    \hat{\theta}_1 = \frac{e^{-|\hat{r}-\frac R 2|^2/\sigma^2}}
    {e^{-|\hat{r} - \frac R 2|^2/\sigma^2} + {e^{-|\hat{r}+\frac R 2|^2/\sigma^2}}},\quad \hat{\theta}_2 = 
    \frac{e^{-|\hat{r} +\frac R 2|^2/\sigma^2}} {e^{-|\hat{r} - \frac R 2|^2/\sigma^2} + {e^{-|\hat{r}+\frac R 2|^2/\sigma^2}}}.
\end{eqnarray}

The final phase-space Hamiltonian  is defined above in Eqs.~\ref{eq:fullHPS}-\ref{eq:theta:used}.  To find the eigenvalues of such a Hamiltonian, we construct  a joint grid of nuclear positions  and  momenta $(R_i, P_j)$; we include $N_R = 100$ regular grid points for nuclear coordinate $R \in \left[2, 4\right] \textup{\AA}$ and $N_P = 100$ grid points for nuclear momentum $P \in \left[-\pi N_P/\Delta R, ..., \pi (N_P-1)/\Delta R\right]$ atomic unit (au) based on the conjugate Fourier transform.
We then take the following steps in in analogy to the BO theory (but now in phase-space):
\mycomment{
\begin{enumerate}
\item[] Loop over $R_i$
\begin{enumerate}
\item[]  Loop over $P_j$
\begin{enumerate}
\item[] Diagonalize the phase-space Hamiltonian (Eq.~\ref{eq:HPS:model}) associated to each joint grid point $\hH_W^{\rm PS}(R_i,P_j) \ket{\psi_0^{\rm PS}} = E^{PS}_{W,0}(R_i,P_j) \ket{\psi_0^{\rm PS}}$. 
\end{enumerate}
\item[] End loop over $P_j$
\end{enumerate}
\begin{enumerate}
\item[] Loop over $R_j$
\begin{enumerate}
\item[] Invert the lowest phase-space eigenenergy $E^{PS}_{W,0}(R_i,P_j)$ to a function in real space $E_0^{\rm PS}(R_i,R_j)$ using Eq.~\ref{eq:weyl}.
\end{enumerate}
\item[] End loop over $R_j$
\end{enumerate}
\item[] End loop over $R_i$
\end{enumerate}
\begin{enumerate}
\item[] Diagonalize the matrix $E_0^{PS}$ (which is a matrix over the nuclear grid in $\bR$ with matrix elements $E_0^{PS}(R_i,R_j)$). 
\end{enumerate}
}


For the  parameter $\sigma$  in Eq.~\ref{eq:theta:used},  we have chosen $\sigma = 1.0$ au.
That being said, the results are not very sensitive to the value of  $\sigma$ in this model problem. For an analysis of the influence of $\sigma$ on our results in this model system, see Appendix \ref{sec:sigma}.

\subsubsection{Harmonic Results}
For the most complete analysis, we will also report results where we consider harmonic BO and phase-space surfaces, i.e. where we force $E_0(R)$ or $E^{\rm PS}_{W,0}(R,P)$ to be a quadratic function.
For the phase-space surface $E^{\rm PS}_{W,0} (R, P)$, we force the energy to be quadratic with respect to both $P$ and $R$ \mycomment{and expand $E^{\rm PS}_{W,0}$ at the minimum energy point $(R_{\rm min}, 0)$: 
\begin{eqnarray} \label{eq:psharmonic1}
E^{\rm PS, harmonic}_{W, 0} =  \frac {\partial^2 E^{\rm PS}_{W,0}}  {\partial P^2}\bigg|_{P = 0} P^2 +  \frac {\partial^2 E^{\rm PS}_{W,0}} {\partial R^2}\bigg|_{R = R_{\rm min}} R^2,
\end{eqnarray}
where the expansion coefficients are calculated by finite difference.
Note that since we have not considered spin-related couplings and there is only one vibrational mode, the phase-space energy will always be minimized at $P = 0$ and the off-diagonal Hessian will be 0, i.e., $\frac {\partial^2 E^{\rm PS}_{W,0}} {\partial R \partial P }= 0$. 
The phase-space harmonic vibrational energy gap is then computed by
\begin{eqnarray} \label{eq:psharmonic2}
    \Delta E = \hbar  \sqrt{\frac {\partial^2 E^{\rm PS}_{W,0}} {\partial R^2} \frac {\partial^2 E^{\rm PS}_{W,0}} {\partial P^2}}.
\end{eqnarray}}

\subsection{Results}

\begin{figure}[ht]
    \centering
    \includegraphics[width=0.5\linewidth]{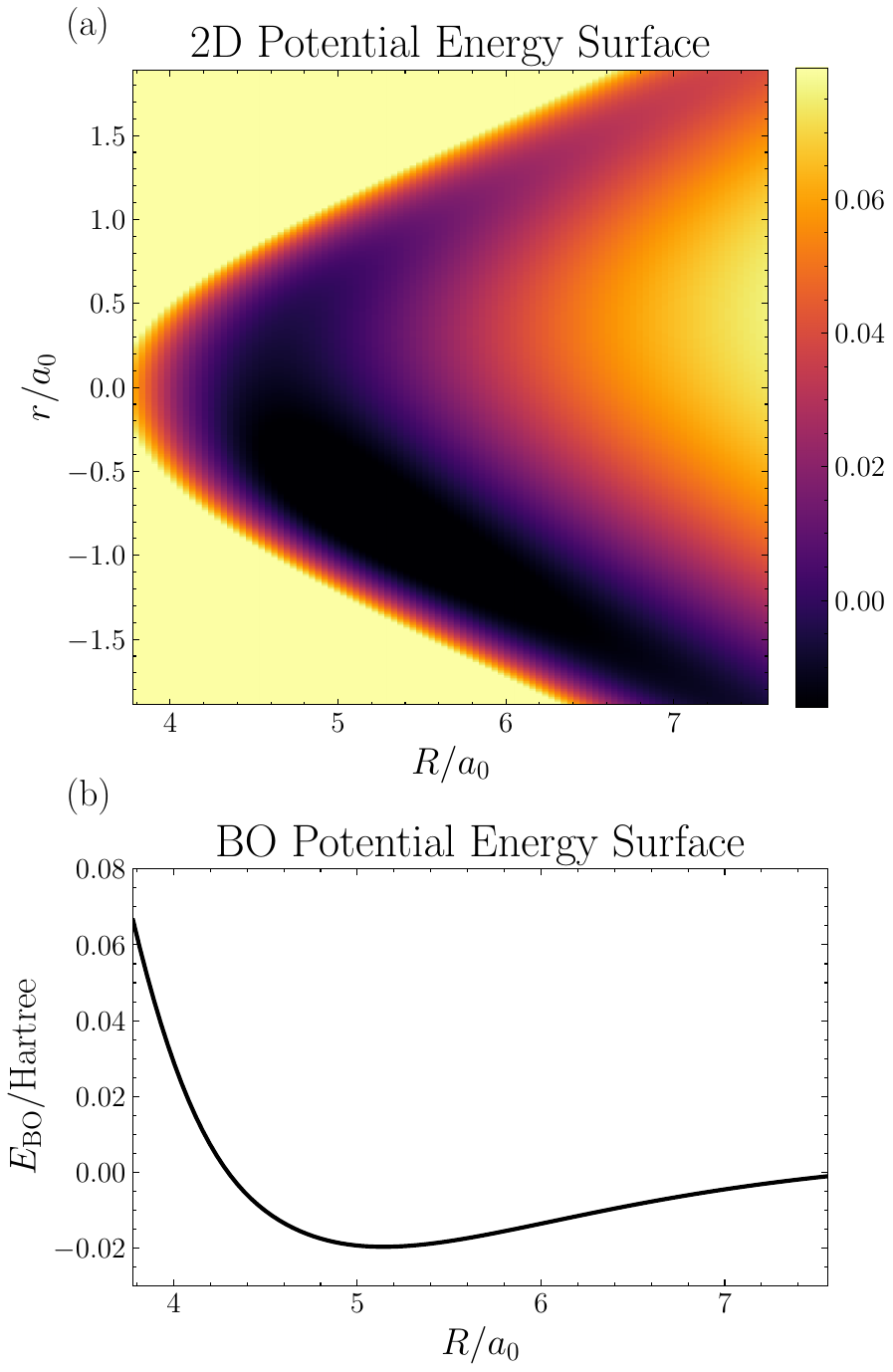}
    \caption{(a) Two-dimensional potential energy surface as a function of inter-nuclear distance ($R$) and distance between electron to the nuclear center of mass ($r$) (b) Morse-like BO potential energy surface obtained from diagonalizing the electronic Hamiltonian as a function of the nuclear position $R$.}
    \label{fig:PES}
\end{figure}

To begin our analysis, in Fig.~\ref{fig:PES}(b),
we plot the ground BO potential energy surface that arises from treating the light and heavy degrees of freedom separately, i.e., treating the oxygen atoms as nuclei and the hydrogen atom as an electron.
One should not expect this BO separation will be a good approximation if the mass ratio $M/m$ is not large enough, such as the actual masses of oxygen and hydrogen atoms (16 and 1 atomic mass units, respectively).  Thus, by varying the nuclei/electron mass ratio $M/m$, we can control non-adiabatic effects for this model Hamiltonian.  
 
\begin{figure}[ht]
    \centering
    \includegraphics[width=0.5\linewidth]{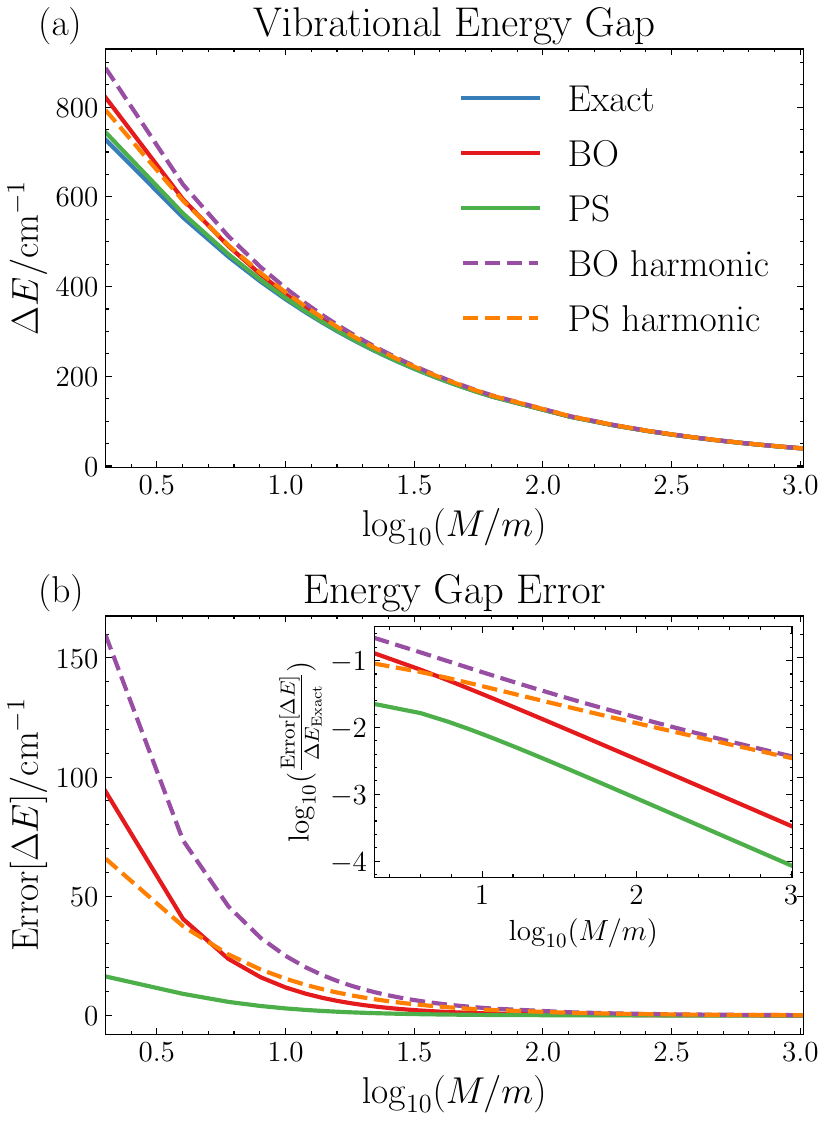}
    \caption{(a) The ground to first excited vibrational energy gap as calculated using various methods. (b) The error in the vibrational energy gap as compared to the exact results. Phase-space methods have far smaller error as compared to BO methods across the entire mass ratio range. (Inset: The relative error on a logarithmic scale.)}
    \label{fig:energy}
\end{figure}

In Fig.~\ref{fig:energy}, we report our results for the ground to first excited vibrational energy gap as a function of the different mass ratio. For all methods, the absolute values of the energy gap decreases as the mass ratio increases and converge to the same results in the large mass ratio limit ($\sim$1000 corresponding to a typical nuclear/electronic mass ratio) in Fig.~\ref{fig:energy}(a). 
Thus,  in Fig.~\ref{fig:energy}(b), we plot the 
error of the energy gap as compared to the exact results. 
In the small mass ratio limit ($\sim$10 corresponding to the oxygen and hydrogen atoms), one can clearly identify that the phase-space method shows a much smaller error compared to the BO method.  
Notably, even within a harmonic approximation, the phase-space method still outperforms the BO and sometimes can be comparable to the quantum BO results. Lastly, to best understand these results, 
the inset of Fig.~\ref{fig:energy}(b) shows the {\em relative} errors in the logarithm scale. As one might expect, the phase-space and BO energy errors both scale with the same power law ($\sim m/M$), but the phase-space method consistently gives an order of magnitude smaller relative error even in the large mass ration limit. 

\begin{figure}[ht]
    \centering
    \includegraphics[width=0.9\linewidth]{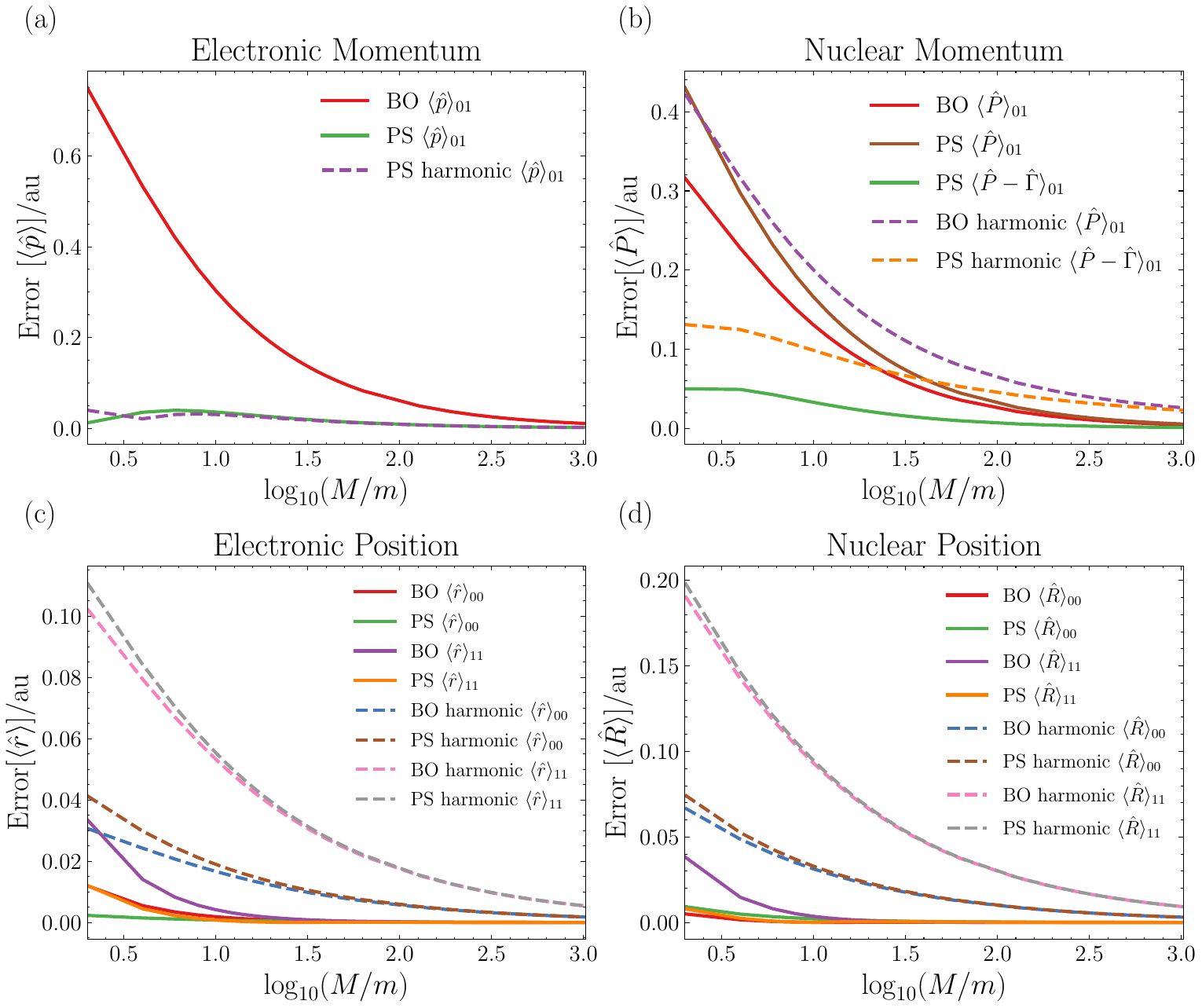}
    \caption{The errors of various observables as calculated from BO and phase-space methods and compared against exact results. We plot the errors in (a) the electronic momentum and (b) the nuclear momentum between the ground and first excited nuclear vibrational state, where the phase-space approach are strong improvements to the BO results. We also plot the errors in (c) electronic position and (d) nuclear position for the ground and first excited vibrational states, where the phase-space and BO results are similar.}
    \label{fig:other}
\end{figure}

It must be emphasized that, with a phase-space Hamiltonian,  one can extract other useful properties as well (besides energy) and these quantities are also more accurate than the corresponding BO Hamiltonian.   For example, in Fig.~\ref{fig:other}, we show the errors in the expectation values of a few nuclear and electronic observables. 
\mycomment{All observables (nuclear or electronic)  for both the exact and BO calculations are calculated on a grid using the final wavefunctions following standard procedures.
For phase-space calculations, nuclear observables $\langle O \rangle_{mn}$ are obtained by evaluating $\langle O \rangle_{mn} = \bra{\chi_m} \hat O \ket{\chi_n}$ where $\chi_m$ and $\chi_n$ are the nuclear (vibrational) wavefunctions as determined from the final step of the flowchart in Sec.~\ref{sec:flowchart}. As for electronic observables, the expectation value $\langle o \rangle$ is calculated by $(i)$ evaluating the electronic observables 
on the phase-space grid $(R_i, P_j)$ with the ground state phase-space electronic wavefunctions $\ket{\psi_0^{\rm PS} (R_i, P_j)}$,
\begin{eqnarray}
    o_W(R_i,P_j) = \bra{\psi_0^{\rm PS}(R_i, P_j)} \hat o \ket{\psi_0^{\rm PS}(R_i, P_j)} 
\end{eqnarray}
and then $(ii)$ averaging over the relevant nuclear Wigner distribution functions (e.g., which are calculated from Eq.~\ref{eq:wigner},   $\rho_{mn}^W (R, P) = \mathcal{W}\left[\ket{\chi_m} \bra{\chi_n} \right]$) :
\begin{eqnarray}
    \langle o \rangle_{mn} = \int dR dP \rho_{mn}^W (R, P)o_W(R,P).
\end{eqnarray}
Let us now evaluate the results.
}

First, for the electronic momentum, it is important to note that BO calculations always predict zero, as $\bra{\psi_k} \hat {\bm p}_{\rm e} \ket{\psi_k} = 0$ for all BO electronic states $\ket{\psi_k}$ at any nuclear geometry. 
However, as shown in Fig.~\ref{fig:other}(a), the phase-space method can capture the electronic momentum between the ground and first excited vibrational states with quantitative accuracy. 
Second, for nuclear momentum, Fig.~\ref{fig:other}(b) demonstrates that phase-space results are also much more accurate than BO methods. 
One interesting nuance is that one must replace the nuclear momentum operator $\hat {\bm P}$ by $\hat {\bm P} - i\hbar\hat {\bm \Gamma}$  in the phase-space formalism which represents a canonical -- kinetic momentum correspondence. 
Overall, for the off-diagonal electronic and nuclear momentum (between the same nuclear vibrational eigenstates), the phase-space method performs very well even with the harmonic approximation. 
As a side note, the on-diagonal nuclear and electronic momentum always vanish due to time reversal symmetry of the whole system (nuclei + electron). 
Finally, in Fig.~\ref{fig:other}(c) and (d), while the nuclear and electronic position observables predicted by BO and phase-space methods are not very different, the phase-space methods do provide slightly better results for the excited vibrational states.   For further analysis, and a decomposition of the importance of different terms in the phase-space Hamiltonian, see Appendix \ref{sec:decomp}. 

\section{Discussion: Implications for Larger Systems }
\mycomment{In this work, we have demonstrated that a novel phase-space view of electronic structure theory can vastly outperform Born-Oppenheimer theory in many facets. Not only can the phase-space approach deliver nonzero and accurate electronic momentum matrix elements between ground state vibrational states (which is impossible for Born-Oppenheimer theory), but the method can also predict molecular vibrational energies  that are far more accurate than those produced by BO theory -- especially when the mass difference between the light and heavy particles is not too large.  

Of course, one might wonder whether the results here can be generalized to larger systems and be of practical importance. After all, Wigner distributions transformations are notoriously expensive to generate. That being said, however, within the harmonic approximation, there is no reason why the present approach should not be immediately practical for very large systems. After all, whereas standard BO theory diagonalizes the Hessian,
\begin{eqnarray}
    \mathbf{H}_{\rm BO} = 
    \left( \frac{\partial E_{\rm BO}}{\partial R_i \partial R_j} \right)
\end{eqnarray}
within phase-space theory, one must  diagonalize (symplectically) a matrix that is only twice as large:
\begin{eqnarray}
    \mathbf{H}_{\rm PS} = \left( \begin{array}{cc}
      \frac{\partial E_{\rm PS}}{\partial R_i \partial R_j} &       \frac{\partial E_{\rm PS}}{\partial R_i \partial P_j} \\
            \frac{\partial E_{\rm PS}}{\partial P_i \partial R_j} &       \frac{\partial E_{\rm PS}}{\partial P_i \partial P_j} 
    \end{array} \right)
\end{eqnarray}
Thus, one can certainly anticipate generating phonons using electronic phase-space Hamiltonians in the immediate future.

In this context, one question that may well arise is how to extract vibrational energies and separate out the center of mass motion, as in Eqs.~\ref{eq:psharmonic1}-\ref{eq:psharmonic2} above. For instance, should one work with internal coordinates as in Eq.~\ref{eq:HPS:model}? Alternatively, one  could set $P_{\rm NCM} = 0$. The latter is size-extensive and likely the most stable. To that end, in Appendix \ref{sec:NCM}, we plot results where we do not throw out the first term in Eq.~\ref{eq:complicated}, but rather set $P_{\rm NCM}$ to zero and diagonalize the Hamiltonian:
\begin{eqnarray} \label{eq:noNCM}
 \hat H_{\rm PS} (R,P) =
   \frac {\left( P- i\hbar \frac{(\hat \Gamma_1 - \hat \Gamma_2)}{2} \right)^2} {2\mu} +  \frac{\hat{p}^2}{2(M+M)} + \frac {\hat p^2} {2m} +  \hat V(R , \hat r).
\end{eqnarray}
From the results in Fig.~\ref{fig:noNCM}, we gather that PS methods still strongly outperforms BO theory -- though the results are not quite as strong as they were for Eq.~\ref{eq:HPS:model}. Future work will need to address the pros and cons of these two internal Hamiltonians. As a side note, we mention that our previous work\cite{duston2024} with VCD spectroscopy used the full Hamiltonian (rather than the internal Hamiltonian without translation).  

Finally, beyond vibrational Hamiltonians, it is worth noting that future work with large systems may well involve excited electronic states, and the present approach should be very useful for so-called phase-space surface hopping dynamics\cite{bian2022,bian2024semiclassical} where we run classical dynamics on phase-space surfaces, with occasional jumps or hops between surface. To that end, it is worth noting that, for the present model, as shown  in Fig.~\ref{fig:excited} in Appendix \ref{sec:excited}, we find that excited state energies are also improved by using phase-space electronic Hamiltonians. Thus, we anticipate that the present phase-space approach will also be very useful for future nonadiabatic calculations based on  classical equations of motion (that do not require a Wigner transform).
}


\section{Future Work and Conclusions}
Looking forward, it would appear clear from the plethora of data collected here and in Refs.\citenum{tao2024,tao2024:bf} that a phase-space approach is more accurate than a BO approach -- and with roughly the same cost. That being said, one must wonder: in what circumstances will the phase-space approach yield meaningfully different answers than BO theory \mycomment{(which performs quite well in most circumstances)}?
To that end, one can imagine several exciting future problems where one would like to go beyond the limitation of BO theory. 
The first potential arena would be to study proton transfer\cite{douhal1996} and proton coupled electron/energy transfer\cite{weinberg2012,pettersson2022} where one could combine a phase-space approach with electronic structure methods that treat both light nuclei (protons) and electrons quantum mechanically (e.g., multi-component methods\cite{kreibich2001,kreibich2008,muolo2020} and nuclear-electronic orbital methods (NEO)\cite{yang2019,hammes2021}). 
Second, it will be fruitful to apply a phase-space approach to study magnetic molecules and solids, where spin-symmetry breaking can lead to intrinsic chiral phonon modes and alter vibrational frequencies\cite{bistoni2021,saparov2022}, and electronic momentum and angular momentum are known to be critically important in such circumstances\cite{schmelcher1988,schmelcher1997}. 
Third, given the {\em ab initio} and semiclassical nature of the phase-space electronic Hamiltonian, one immediate goal is to investigate nuclear (vibrational) effects at curve crossings during electron transfer processes\cite{bian2021} and decipher whether phase-space approaches (that include angular momentum conservation\cite{bian2023}) can help explain the nature of nuclear effects in chiral-induced spin selectivity\cite{zollner2020,das2022,bloom2024}. 
Fourth, the present  phase-space approach opens new and inexpensive avenues for exploring electron-phonon interactions, which could lead to a better atomistic understanding of superconductivity\cite{frohlich1954,pietronero1995,lanzara2001}.   
At the end of the day, there is a very real chance that phase-space electronic Hamiltonian approaches are here to stay and that a new world of very novel chemistry awaits us.

\section{Acknowledgements}
This work was supported by the U.S. Air Force Office of Scientific Research (AFOSR) under Grant No. FA9550-23-1-0368 and No. FA9550-18-1-420 (J.E.S.).

\appendix
\section{Influence of the $\sigma$ parameter in our model problem} \label{sec:sigma}
\begin{figure}[ht]
    \centering
    \includegraphics[width=0.6\linewidth]{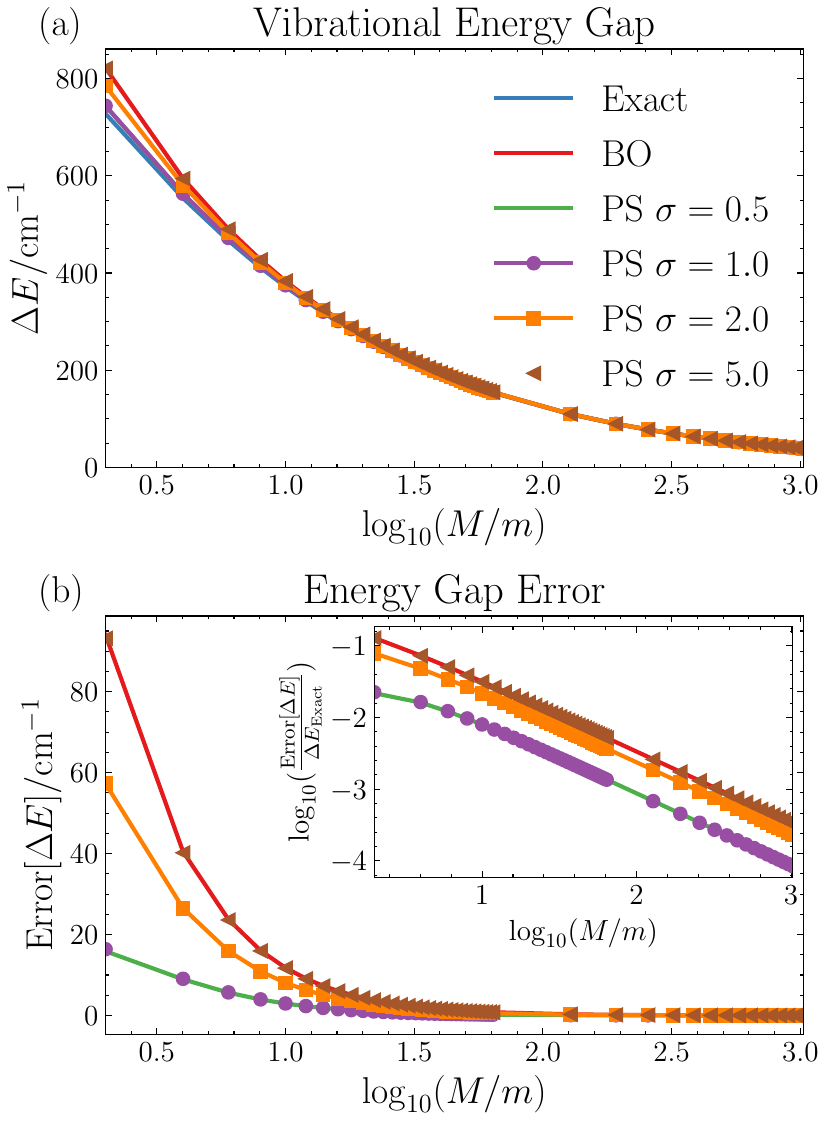}
    \caption{(a) The ground to first excited vibrational energy gap and (b) the errors in that gap as calculated  for BO theory and  phase-space methods (i.e. with different  $\sigma$ values).}
    \label{fig:sigma}
\end{figure}

As shown in Fig.~\ref{fig:sigma}, our results are effectively unchanged for small width values, $\sigma = 0.5$ au and $\sigma = 1.0$ au, which represent the localized $\hat \Gamma$ limit; in both cases, these results vastly outperform BO. For large $\sigma$ values (where we do not enforce strong locality), e.g., $\sigma = 5.0$ au, the phase-space methods simply recover the BO results.

\section{Decomposition of the phase-space electronic Hamiltonian}
\label{sec:decomp}
Finally, to gain a better understanding of the phase-space Hamiltonian in vibrational problems, we can decompose the phase-space correction terms in Eq.~\ref{eq:hps:general} into two parts: a $\hat {\bm \Gamma} \cdot \bm P$ contribution and a $\hat {\bm \Gamma} \cdot \hat {\bm \Gamma}$ contribution. Then, we performed phase-space calculations for which only one of these two terms is included in the Hamiltonian. 

\begin{figure}[ht]
    \centering
    \includegraphics[width=0.6\linewidth]{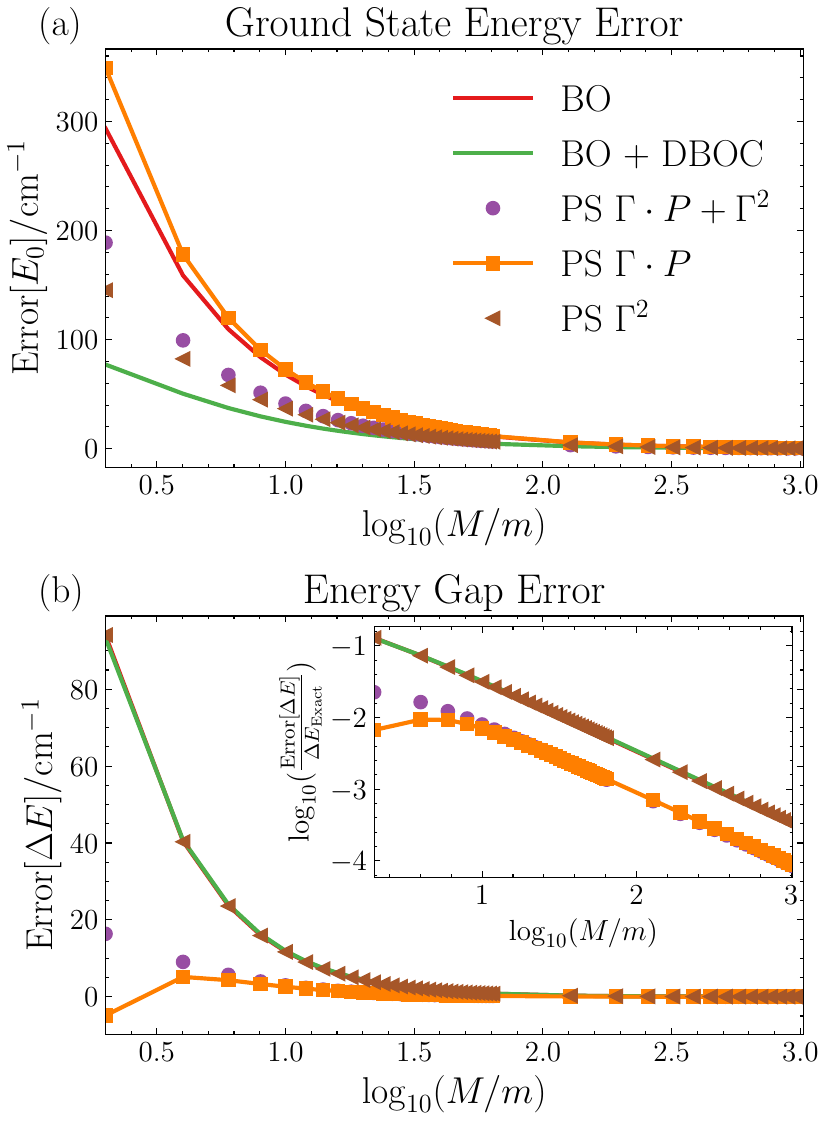}
    \caption{Errors of (a) the ground state absolute energy and (b) ground to first excited vibrational energy gap for (1) BO (2), BO+DBOC and (3) phase-space calculations (with different contributions) as  compared against  exact results.}
    \label{fig:decompose}
\end{figure}

In Fig.~\ref{fig:decompose}(a), we present the errors in the energies for the ground vibrational state  ($E_0$) as calculated using different methods; to be clear, here we do not consider the error in the energy gap  ($E_1-E_0$) but rather the error of the ground state alone ($E_0$).
Overall,  we find that both (i) the complete phase-space results (using $\hat {\bm \Gamma} \cdot \hat {\bm \Gamma}$ and $\hat {\bm \Gamma} \cdot \bm P$ from Eq.~\ref{eq:HPS}, purple lines)  and (ii) the partial phase-space results using  $\hat {\bm \Gamma} \cdot \hat {\bm \Gamma}$ (triangles) approach the BO plus diagonal Born-Oppenheimer correction (DBOC) results.
Obviously, as far as the absolute ground state energy is concerned, the  $\hat {\bm \Gamma} \cdot \hat {\bm \Gamma}$ term is the most important correction. 


Next, we turn to the energy gap, $E_1 - E_0$.
In Fig.~\ref{fig:decompose}(a), we plot the errors in the energy gap for the same methods. 
Here, we find that neither $\hat {\bm \Gamma} \cdot \hat {\bm \Gamma}$ nor BO+DBOC offer any improvement.  For these problems, we clearly see that including a  momentum-dependent potential $\hat {\bm \Gamma} \cdot \bm P$ is critical for recovering an accurate energy gap.  In a sense, these results should be intuitive.
After all, in the hydrogen atom case, $\hat {\bm \Gamma} \cdot \bm P$ and $\hat {\bm \Gamma} \cdot \hat{\bm \Gamma}$ correspond exactly to  adiabatic (i.e., DBOC) and non-adiabatic corrections to  BO theory as discussed in Refs.~\cite{moss1996,kutzelnigg2007}.  Furthermore, similar results as far the effect of the DBOC on absolute and relative energies were reported by Hammes-Schiffer {\em et al.}\cite{schneider2019}

\section{Results without eliminating the MCM motion} \label{sec:NCM}
\mycomment{We have performed the same calculation using a phase-space Hamiltonian (Eq.~\ref{eq:noNCM}) that does not eliminate the MCM motion but instead imposes the condition  $P_{\rm NCM} = 0$.
Both phase-space results, calculated from Eq. \ref{eq:HPS:model} and Eq.~\ref{eq:noNCM}, outperform the BO results, with the removal of MCM motion yielding better accuracy. For large systems, with large total mass $M$, these two methods should be very similar. That being said, formally speaking, Eq. \ref{eq:hps:general} is size consistent (meaning that the vibrational energies for a molecule will be the same if we one or many noninteracting molecules)-- but this size consistency is eliminated if we remove the total translational energy as in Eq.~\ref{eq:noNCM} and set $\tilde{H}_{PS} = \hat{H}_{PS} - \frac{\hat{p}^2}{2\sum_A M_A}$.  For this reason, the results using Eq. \ref{eq:HPS:model} are more likely to be relevant when investigating large systems. }
\begin{figure}[H]
    \centering
    \includegraphics[width=0.6\linewidth]{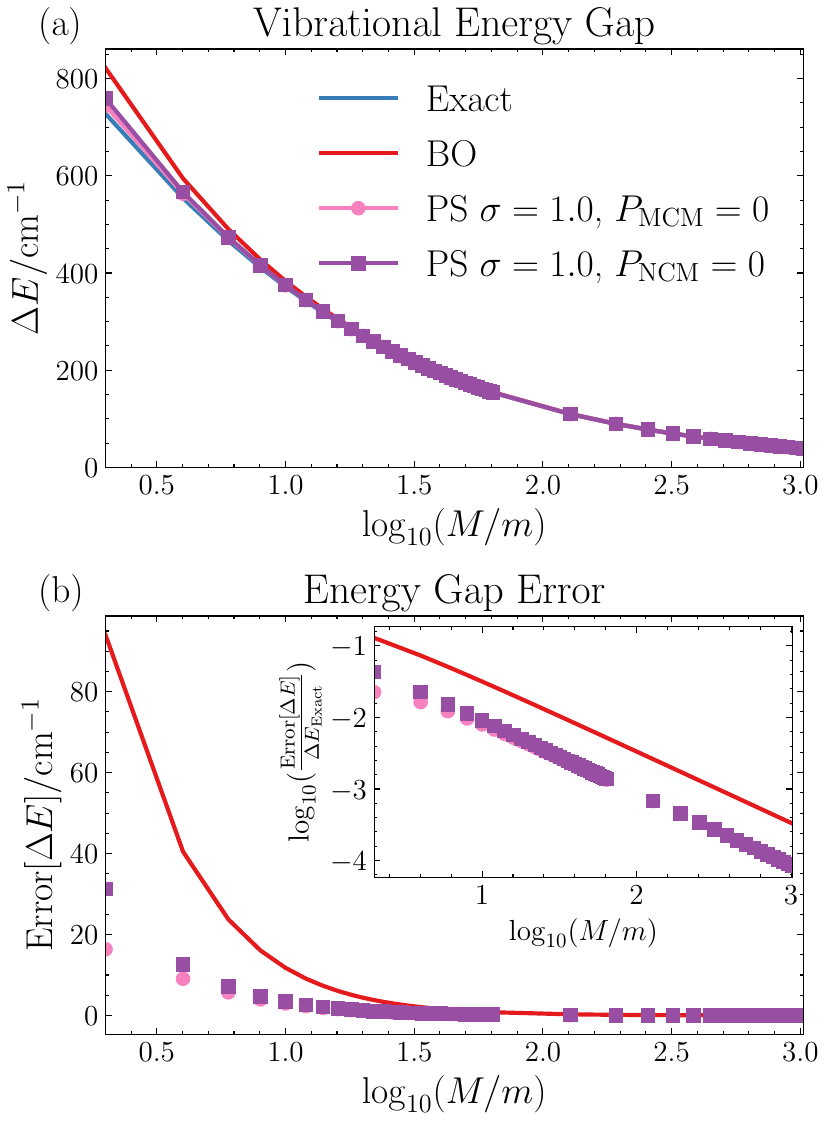}
    \caption{(a) The ground to first excited vibrational energy gap and (b) the errors in that gap as calculated  for BO theory and phase-space methods (pink circle with Eq.~\ref{eq:HPS:model} and purple box with Eq.~\ref{eq:noNCM}). Both phase space approaches strongly outperform BO.}
    \label{fig:noNCM}
\end{figure}

\section{Vibrational energies calculated for both ground and excited electronic surfaces} \label{sec:excited}
We have also calculated the errors in the vibrational energy gaps for a mass ratio of  $M/m=16$ in Eq.~\ref{eq:HPS:model}. The phase-space methods show improvement in the vibrational energies on both the ground and excited electronic energy surfaces, highlighting another area where phase-space approaches may well find application.
\begin{figure}[H]
    \centering
    \includegraphics[width=0.6\linewidth]{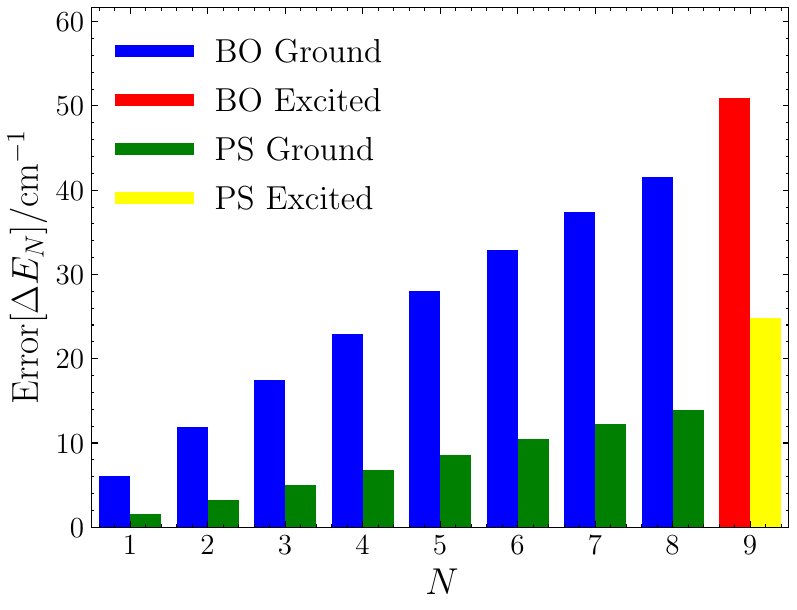}
    \caption{The errors in the vibrational energy gaps $\Delta E_N = E_N - E_0$ are calculated for both BO theory and phase-space methods.  The vibrational states are ordered by their energies, with $E_N$ representing the energy of the $N$-th vibrational states irrespective of whether the latter corresponds to a ground or excited electronic surface. Vibrational states on the ground BO electronic surface are labeled in blue (BO) and green (PS), while those on the excited electronic surface are labeled in red (BO) and yellow (PS). Note that phase space approaches outperform BO both for ground and excited state vibrational energies.}
    \label{fig:excited}
\end{figure}

\pagebreak
\pagebreak

\bibliography{main}
\end{document}